# An improved PINN framework integrating localized collocation scheme and PIKF

Qiang Xi[a,b], Wenzhi Xu[b], Mario Cvetković [c], Dragan Poljak[c], Timon Rabczuk [d], Zhuojia Fu[a,b,1],

[a]Key Laboratory of Ministry of Education for Coastal Disaster and Protection, Hohai University, Nanjing 210098, China
[b]Center for Numerical Simulation Software in Engineering and Sciences, College of Mechanics and Engineering Science, Hohai University, Nanjing 211100, China
[c]Faculty of Electrical Engineering, Mechanical Engineering and Naval Architecture, University of Split, Rudera Boskovica 32, 21000 Split, Croatia
[d]Institute of Structural Mechanics, Bauhaus-University Weimar, Weimar 99423, Germany

**Abstract:** We propose a localized physics-informed kernel function neural network (LPIKFNN), which is an improved physics-informed neural network (PINN) based on physics-informed kernel function (PIKF). In the LPIKFNN framework, the localized collocation scheme discretizes the physical quantities within the local domain, where the physical field is represented as a linear combination of PIKFs. Based on this representation, the multilayer perceptron is trained to iteratively learn the physical quantities. To overcome the computational challenges of conventional PINN in higher-order derivative and high wavenumber problems, the LPIKFNN constructs the loss function using the PIKF and a localized collocation scheme rather than relying on automatic differentiation. As a result, the costly derivative evaluations required to enforce governing equations during iterative training are eliminated, leading to significantly improved computational efficiency and training performance. Moreover, incorporating PIKFs into the loss function enables the proposed LPIKFNN to significantly improve computational accuracy in high-wavenumber problems characterized by highly oscillatory physical fields. To overcome the computational bottleneck of the physics-informed kernel function neural network (PIKFNN) in heterogeneous problems, the LPIKFNN introduces a localized collocation scheme that removes reliance on global PIKFs, enabling accurate predictions where global PIKFs are unavailable. The feasibility and accuracy of the proposed LPIKFNN are demonstrated through a series of benchmark studies, including high wavenumber problems, higher-order derivative problems, nonlinear problems, heterogeneous problems, and potential-based inverse electromyography. The numerical predictions obtained by LPIKFNN show excellent agreement with available analytical solutions and experimental measurements.

---

[1]Corresponding authors: paul212063@hhu.edu.cn

## 1. Introduction

Partial differential equations (PDEs) play a fundamental role in modeling and analyzing a wide range of physical phenomena in science and engineering. They govern both forward problems, where unknown physical fields within a computational domain are predicted from prescribed boundary conditions, and inverse Cauchy problems, where unknown boundary quantities are inferred from available observations. Consequently, the development of efficient and accurate numerical methods for solving PDEs is essential for reliable simulation and prediction of complex physical phenomena in engineering applications.

A wide range of numerical methods has been developed for solving PDEs. Among them, classical mesh-based methods, including the finite element method [1,2], finite difference method [3,4], finite volume method [5,6], and boundary element method [7,8], have long served as the dominant computational tools in engineering analysis. However, the performance of these methods may deteriorate in the presence of severe mesh distortion, and the generation of high-quality meshes for complex geometries can be computationally expensive. To address these challenges, significant research efforts have been devoted to the development of meshless methods. Representative approaches, such as smoothed particle hydrodynamics [9,10], the element-free Galerkin method [11,12], the radial basis function collocation method [13,14], and the generalized finite difference method [15,16], have demonstrated promising performance for solving PDEs in fluid mechanics, fracture mechanics, and elastic mechanics.

Recently, physics-informed neural networks (PINNs) [17-21] that integrate physical laws with data-driven learning have been proposed as an effective method for solving PDEs. Only node information is required in the prediction of PINNs, which fundamentally avoids the mesh dependency problem and can also be regarded as a meshless method. By incorporating the governing PDEs and boundary conditions into the loss function and embedding them within the training process of neural networks, PINNs substantially reduce the reliance of conventional data-driven neural networks on large datasets [22]. In particular, it has demonstrated strong capabilities in addressing high-dimensional problems, complex boundary conditions, and inverse problems [23,24]. Furthermore, the development of numerous open-source libraries [25-27] has significantly lowered the barrier to adoption of PINNs in scientific and engineering applications.

The most significant innovation of PINNs is the integration of governing PDEs into the loss function of neural networks. In this framework, automatic differentiation plays a central role by simplifying the derivation of PDE residuals and their derivatives. However, as the complexity and dimensionality of the PDEs increase, the computational cost of automatic differentiation grows substantially [28]. Moreover, automatic differentiation can exhibit numerical instability when calculating higher-order derivatives or handling high-wavenumber solutions, which may compromise prediction accuracy or even produce incorrect results. Consequently, considerable research efforts have been devoted to developing alternative strategies [29] and novel neural network architectures [30-33]. For instance, boundary-integrated neural networks replace automatic differentiation with the boundary integral equation in constructing PDE loss functions and have been successfully applied to potential problems [34], acoustic radiation and scattering [35], elastostatics and piezoelectricity [36]. Zhong et al. [37] utilized interpolation functions and implicit Runge-Kutta techniques to construct the loss function for low-temperature plasma equations, thus proposing the Runge-Kutta physics-informed neural network. Physics-informed kernel function neural network (PIKFNN) [38,39] replaces the activation function with the physics-informed kernel function (PIKF) to construct the loss function, and achieves efficient computation of various linear and nonlinear equations.

Inspired by the PIKFNN and localized method of fundamental solutions (LMFS) [40,41], the PIKF and localized collocation scheme are employed to construct the loss function in place of automatic differentiation. Building on this idea, we propose a localized physics-informed kernel function neural network (LPIKFNN). The proposed framework directly implements a localized discretization of governing PDEs and boundary conditions, where the physical field within the local domain is represented as a linear combination of PIKFs [42,43], substantially reducing model training time. By incorporating PIKFs that capture the intrinsic variation of the PDE solutions, LPIKFNN achieves superior accuracy in predicting highly fluctuating physical phenomena. Furthermore, the introduction of a localized collocation scheme removes the reliance on global PIKFs in the PIKFNN, thereby enabling accurate predictions for heterogeneous problems where a global PIKF is unavailable.

The remainder of this paper is organized as follows. Section 2 outlines the theoretical foundations and computational frameworks of the LPIKFNN. Subsequently, six numerical examples are presented in Section 3 to demonstrate the accuracy and efficiency of the proposed LPIKFNN in predicting high wavenumber problems, higher-order derivative problems, nonlinear problems, heterogeneous problems, and potential-based inverse electromyography. Finally, some key conclusions and remarks of this study are presented in Section 4.

## 2. Computational frameworks of LPIKFNN

Without loss of generality, the governing equations and boundary conditions of the steady-state physics problem are given by

$$\mathbf{L}u(\mathbf{x}) = 0, \mathbf{x} \in \Omega, \tag{1}$$

$$\mathbf{B}u(\mathbf{x}) = g(\mathbf{x}), \mathbf{x} \in \Gamma, \tag{2}$$

where $u(\mathbf{x})$ is the physical quantity (such as sound pressure, temperature, electrical potential etc.) at the spatial coordinate $\mathbf{x}$, $\mathbf{L}$ represents the linear or nonlinear partial differential operator, $\mathbf{B}$ indicates the boundary differential operator, $g(\mathbf{x})$ denotes the boundary condition function, $\Omega$ and $\Gamma$ indicate the computational domain and boundary. The proposed LPIKFNN integrates the multilayer perceptron-based machine learning approach, physics-informed kernel functions, and a localized collocation scheme to predict solutions of the steady-state physics problem.

### 2.1 Multilayer perceptron machine learning framework

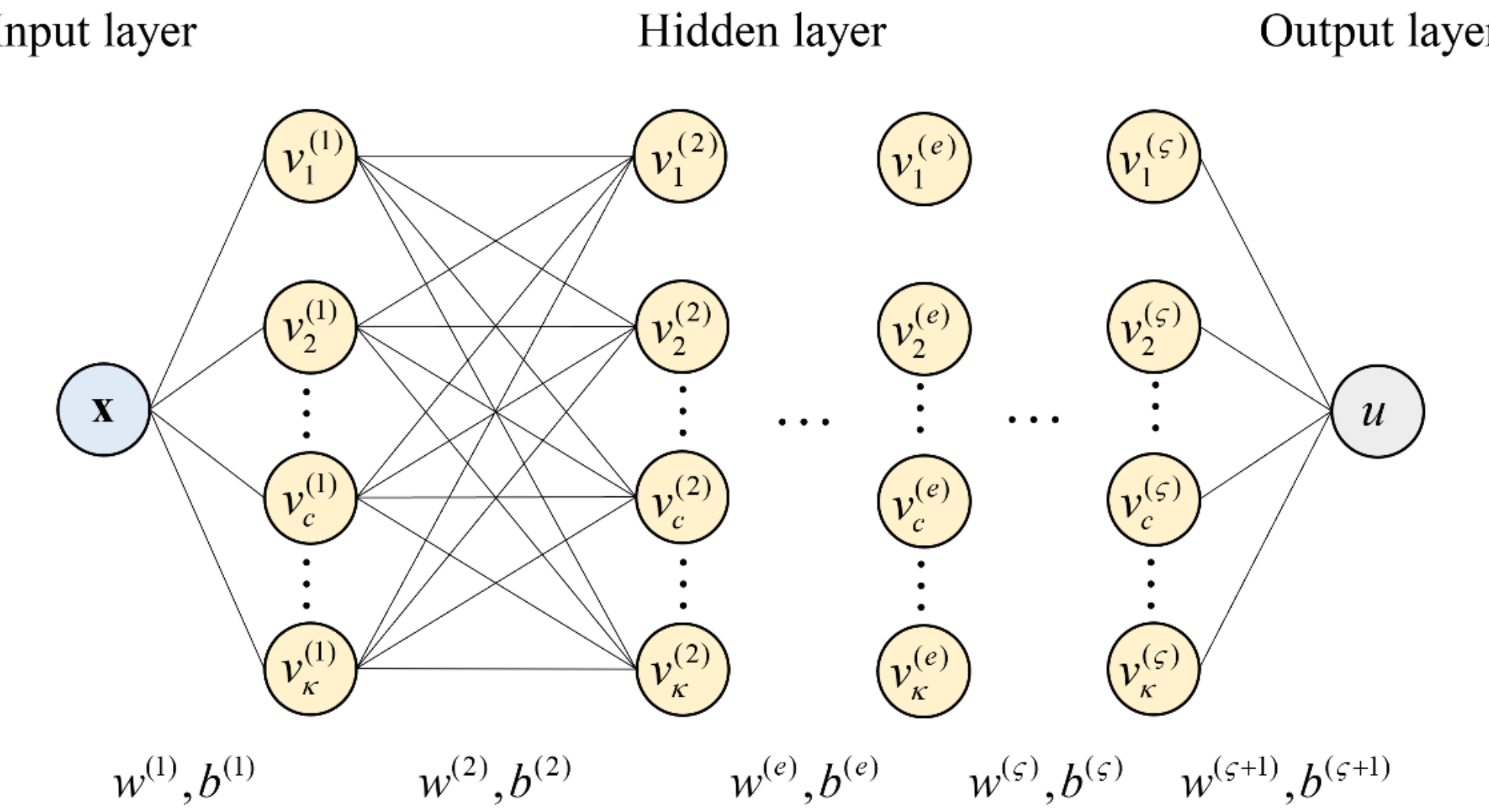

Fig. 1. Schematic diagram of the multilayer perceptron machine learning framework.

The multilayer perceptron component forms a fundamental part of the LPIKFNN and consists of three layers: input, hidden, and output, as illustrated in Fig. 1. The spatial coordinates of the training nodes serve as input variables, which are then processed through a series of linear and nonlinear transformations to produce the output physical quantities. Specifically, the linear transformation computes a weighted sum of the previous layer's neuron values plus a bias term, while the nonlinear transformation is introduced through the activation function to capture the network's nonlinearity.

The predicted values of neurons in the hidden layer can be represented as

$$v_c^{(e)} = \eta\left(\mathbf{W}^{(e)}\mathbf{v}^{(e-1)} + b^{(e)}\right), \tag{3}$$

where $v_c^{(e)}$ is the predicted value of the $c$th neuron in the $e$th hidden layer, $\mathbf{W}^{(e)}$ represents the weight vector of the $e$th hidden layer, $b^{(e)}$ denotes the bias of the $e$th hidden layer, $\eta$ is the activation function, $\mathbf{v}^{(e-1)}$ indicates the predicted value vector of neurons in the $(e-1)$th hidden layer, and $\mathbf{v}^{(0)}$ denotes the spatial coordinate $\mathbf{x}$ of the training node.

After the data propagates from the input layer to the last hidden layer, the predicted physical quantity generated by the multilayer perceptron is expressed as

$$u(\mathbf{x},\theta) = \eta\left(\mathbf{W}^{(\varsigma+1)}\mathbf{v}^{(\varsigma)} + b^{(\varsigma+1)}\right), \tag{4}$$

where $\varsigma$ represents the total number of hidden layers, $u(\mathbf{x},\theta)$ indicates the predicted physical quantity at the spatial coordinate $\mathbf{x}$, $\theta$ denotes the hyperparameters of the neural network, comprising the weights and biases. Notably, these hyperparameters are iteratively optimized during training to minimize the loss function.

### 2.2 Loss function constructed by PIKF and localized collocation scheme

The loss function is a critical factor in determining the prediction accuracy of the LPIKFNN. For steady-state physics problems, the total loss function $\Xi_{total}$ typically comprises the PDE residual loss function $\Xi_{pde}$, the boundary condition loss function $\Xi_{bc}$, and the data loss function $\Xi_{data}$. In PINN predictions, automatic differentiation is used to compute the partial differential terms of the predicted solution, which are then substituted into the governing equations and boundary conditions to form the loss function. In contrast, the LPIKFNN employs the PIKF and localized collocation scheme to construct the loss function without relying on automatic differentiation.

#### 2.2.1 Forward problem framework

In LPIKFNN predictions, the PIKF satisfying the characteristics of the governing equation in the local domain is constructed, avoiding the dependence of PIKFNN on the global PIKF, thus overcoming the derivation difficulties of the global PIKF in heterogeneous and nonlinear problems. The physical quantity of the training node within the local domain is represented as a linear combination of PIKFs, one can obtain

$$u\left(\mathbf{x}^{(l)}\right)=\sum_{j=1}^{M}\alpha_j^{(l)}F\left(\mathbf{x}^{(l)},\mathbf{s}_j^{(l)}\right)=\mathbf{f}\boldsymbol{\alpha}_f, \tag{5}$$

where $\mathbf{x}^{(l)}$ is the training nodes, $\mathbf{s}_j^{(k)}$ represents the source nodes, $M$ is the number of source nodes, $\boldsymbol{\alpha}_f$ indicates the coefficient vector, $\mathbf{f}$ is the vector formed by the PIKF $F\left(\mathbf{x}^{(l)},\mathbf{s}_j^{(l)}\right)$. The local domain is shown in Fig. 2.

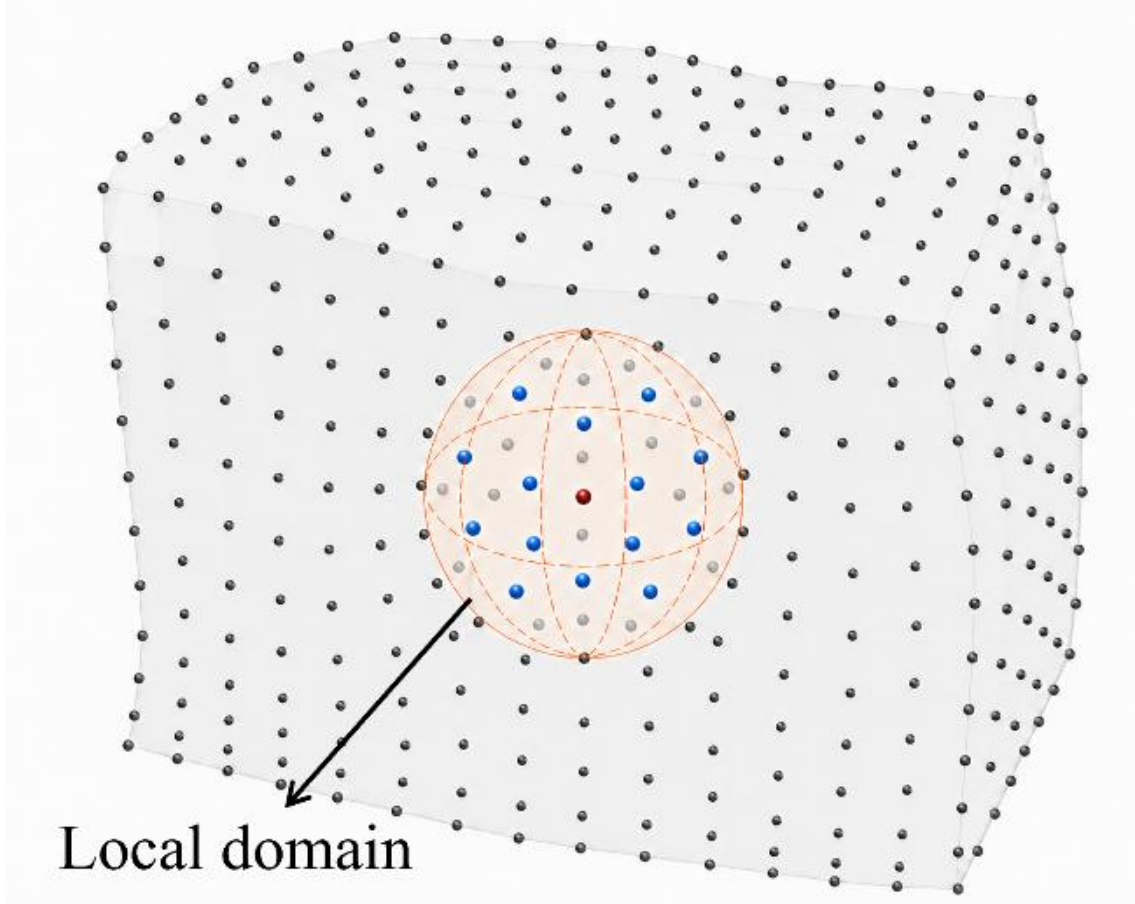

Fig. 2. Schematic plot of training nodes and local domain in LPIKFNN.

By using Eq. (5) to calculate all the training nods in the local domain, it can be rewritten in the following matrix form

$$\mathbf{u}=\mathbf{F}\boldsymbol{\alpha}_f, \tag{6}$$

where $\mathbf{u}$ represents the column vector formed by the physical quantities of the training nodes in the local domain, $\mathbf{F}$ indicates the matrix formed by the PIKF. The coefficient vector $\boldsymbol{\alpha}_f$ can be represented as

$$\boldsymbol{\alpha}_f=\mathbf{F}^{-1}\mathbf{u}. \tag{7}$$

Then, Substituting the coefficient vector $\boldsymbol{\alpha}_f$ calculated in Eq. (7) into Eq. (5), one can obtain

$$u\left(\mathbf{x}^{(l)}\right)=\mathbf{f}\mathbf{F}^{-1}\mathbf{u}=\sum_{j=1}^{M}\omega_j^{(f)}u\left(\mathbf{x}_j^{(l)}\right), \tag{8}$$

where $\omega_j^{(f)}$ indicates the weighing coefficients in the discretization of the governing equation, which indicates the relationship of physical quantities at each training node in the local domain. Next substituting the $u\left(\mathbf{x}^{(l)}\right)$ obtained from Eq. (8) into the boundary condition (2) will yield

$$g\left(\mathbf{x}^{(l)}\right)=\mathbf{B}u\left(\mathbf{x}^{(l)}\right)=\sum_{j=1}^{M}\mathbf{B}\omega_j^{(f)}u_j=\sum_{j=1}^{M}\varpi_j^{(f)}u\left(\mathbf{x}_j^{(l)}\right), \tag{9}$$

where $\varpi_j^{(f)}$ represents the weighing coefficients in the discretization of the boundary condition.

Based on Eqs. (8) and (9), the PDE residual loss function and the boundary condition loss function are defined by

$$\Xi_{pde}\sum_{l=1}^{N_{pde}}\left|u\left(\hat{\mathbf{x}}^{(l)}\right)-\sum_{j=1}^{M}\omega_j^{(f)}u\left(\hat{\mathbf{x}}_j^{(l)}\right)\right|^2, \tag{10}$$

$$\Xi_{bc}=\sum_{l=1}^{N_{bc}}\left|g\left(\overline{\mathbf{x}}^{(l)}\right)-\sum_{j=1}^{M}\varpi_j^{(f)}u\left(\overline{\mathbf{x}}_j^{(l)}\right)\right|^2, \tag{11}$$

where $N_{pde}$ is the number of training nodes $\hat{\mathbf{x}}^{(l)}$ that satisfy the PDE, $N_{bc}$ indicates the number of boundary training nodes $\overline{\mathbf{x}}^{(l)}$. Meanwhile, the data loss function enforces data consistency by minimizing the difference between the predicted values and the experimental data, which is given by

$$\Xi_{data}=\sum_{l=1}^{N_{data}}\left|u\left(\tilde{\mathbf{x}}^{(l)}\right)-h\left(\tilde{\mathbf{x}}^{(l)}\right)\right|^2, \tag{12}$$

where $N_{data}$ is the number of training nodes $\tilde{\mathbf{x}}^{(l)}$ with available experimental data $h\left(\tilde{\mathbf{x}}^{(l)}\right)$. Finally, the total loss function is obtained by a linear combination of the PDE residual loss function, the boundary condition loss function and the data loss function as

$$\Xi_{total}\left(\theta\right)=\frac{\Xi_{pde}+\Xi_{bc}+\Xi_{data}}{N_{pde}+N_{bc}+N_{data}}. \tag{13}$$

After defining the total loss function of the LPIKFNN, the L-BFGS optimizer is used to optimize the network hyperparameters $\theta$ to minimize this total loss function. The schematic of the LPIKFNN forward problem framework is illustrated in Fig. 3.

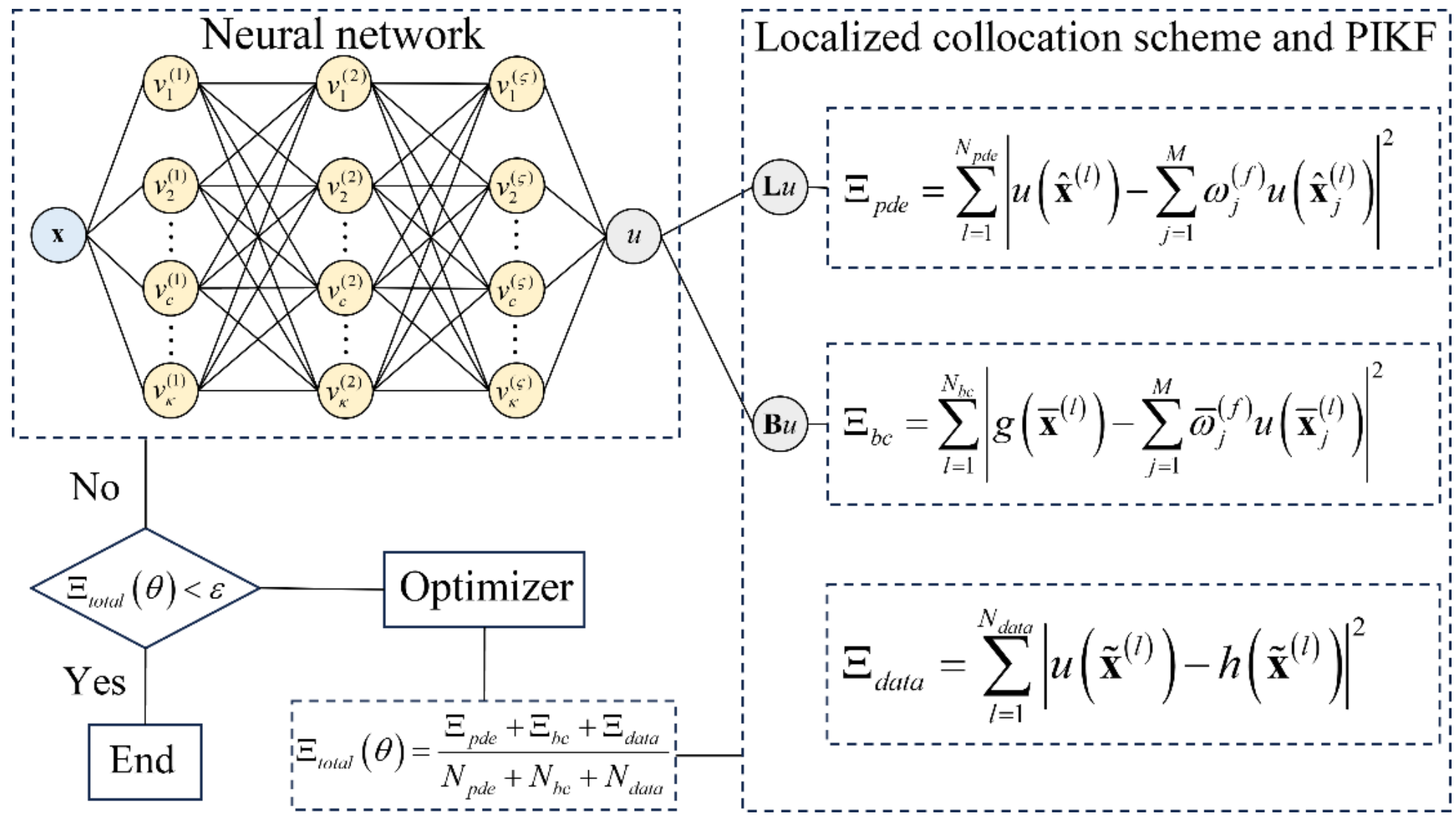

Fig. 3. Schematic diagram of the LPIKFNN forward problem framework.

It is noteworthy that the weighting coefficients in the PDE residual and boundary condition loss functions depend solely on the PIKFs and the coordinates of the training nodes. Consequently, the weight functions do not need to be recomputed during LPIKFNN iterations, which substantially reduces training time. In addition, because the PIKFs inherently capture the variation characteristics of the governing equations, the LPIKFNN can accurately and efficiently resolve problems with highly fluctuating physical quantities, such as high-wavenumber phenomena. Moreover, the introduction of a localized collocation scheme removes the reliance on global PIKFs in the PIKFNN, thereby enabling accurate predictions for heterogeneous problems where a global PIKF is unavailable.

### 2.2.2 Inverse Cauchy problem framework

Based on the LPIKFNN framework for forward problems introduced in the previous subsection, this section further presents the LPIKFNN framework for the inverse Cauchy problem. The governing equations of the inverse Cauchy problem of steady-state physical fields remain as expressed in Eq. (1), while two different boundary conditions are typically imposed on the measurable boundary, which can be prescribed as

$$\mathbf{B}_1 u(\mathbf{x}) = g_1(\mathbf{x}), \mathbf{x} \in \Gamma_m, \tag{14}$$

$$\mathbf{B}_2 u(\mathbf{x}) = g_2(\mathbf{x}), \mathbf{x} \in \Gamma_m, \tag{15}$$

where $\mathbf{B}_1$ and $\mathbf{B}_2$ represent two different boundary differential operators, $g_1(\mathbf{x})$

and $g_2\left(\mathbf{x}\right)$ indicate two different boundary condition function, $\Gamma_m$ is the measurable boundary.

The essence of the inverse Cauchy problem is to infer unknown physical quantities on the unmeasurable boundary from a limited number of detection data on the measurable boundary. The key distinction between the inverse Cauchy and the forward problem framework is that the former requires incorporating the unknown physical quantities as part of the loss function into the optimization process.

In the numerical implementation, the training nodes associated with the unknown physical quantities are denoted as $\hat{\mathbf{x}}^{(u)}$, while the original training nodes for the governing equation are denoted as $\hat{\mathbf{x}}^{(o)}$. To enforce that the unknown physical quantities satisfy the governing equation, they are incorporated into the PDE loss function (10), which can be rewritten as

$$\Xi_{pde}^{(ic)} = \sum_{l=1}^{N_{pde}^{(ic)}} \left| u\left(\hat{\mathbf{x}}^{(l)}\right) - \sum_{j=1}^{M} \omega_j^{(f)} u\left(\hat{\mathbf{x}}_j^{(l)}\right) \right|^2, \tag{16}$$

where $\hat{\mathbf{x}}^{(l)} = [\hat{\mathbf{x}}^{(u)}, \hat{\mathbf{x}}^{(o)}]$, $N_{pde}^{(u)}$ and $N_{pde}^{(o)}$ represent the number of training nodes $\hat{\mathbf{x}}^{(u)}$ and $\hat{\mathbf{x}}^{(o)}$, $N_{pde}^{(ic)} = N_{pde}^{(u)} + N_{pde}^{(o)}$. Meanwhile, the boundary condition loss function enforces boundary consistency constraints by minimizing the error between different boundary functions and predicted boundary values, which can be prescribed as

$$\Xi_{bc}^{(ic)} = \Xi_{bc1}^{(ic)} + \Xi_{bc2}^{(ic)}, \tag{17}$$

in which

$$\Xi_{bc1}^{(ic)} = \sum_{l=1}^{N_{bc}^{(ic)}} \left| g_1\left(\overline{\mathbf{x}}^{(l)}\right) - \sum_{j=1}^{M} \widehat{\omega}_j^{(f)} u\left(\overline{\mathbf{x}}_j^{(l)}\right) \right|^2, \tag{18}$$

$$\Xi_{bc2}^{(ic)} = \sum_{l=1}^{N_{bc}^{(ic)}} \left| g_2\left(\overline{\mathbf{x}}^{(l)}\right) - \sum_{j=1}^{M} \breve{\omega}_j^{(f)} u\left(\overline{\mathbf{x}}_j^{(l)}\right) \right|^2, \tag{19}$$

where $N_{bc}^{(ic)}$ is the number of training nodes $\overline{\mathbf{x}}^{(l)}$ on the measurable boundary. Furthermore, the boundary condition loss function and data loss function in the inverse Cauchy problem framework are given by

$$\Xi_{data}^{(ic)} = \sum_{l=1}^{N_{data}^{(ic)}} \left| u\left(\tilde{\mathbf{x}}^{(l)}\right) - h\left(\tilde{\mathbf{x}}^{(l)}\right) \right|^2, \tag{20}$$

$$\Xi_{total}^{(ic)}\left(\theta\right)=\frac{\Xi_{pde}^{(ic)}+\Xi_{bc}^{(ic)}+\Xi_{data}^{(ic)}}{N_{pde}^{(ic)}+2N_{bc}^{(ic)}+N_{data}^{(ic)}}, \tag{21}$$

where $N_{data}^{(ic)}$ is the number of training nodes $\tilde{\mathbf{x}}^{(l)}$ with available experimental data.

Finally, the L-BFGS optimizer is applied to minimize the total loss function. It should be noted that only the network weights and biases are optimized in the forward problem framework, whereas the optimization simultaneously includes the network weights, biases, and the unknown physical quantities in the inverse Cauchy problem framework. In summary, the primary modification in the LPIKFNN inverse Cauchy framework is the incorporation of unknown physical quantities into both the loss function and the optimization process. The inverse Cauchy problem framework is illustrated in Fig. 4.

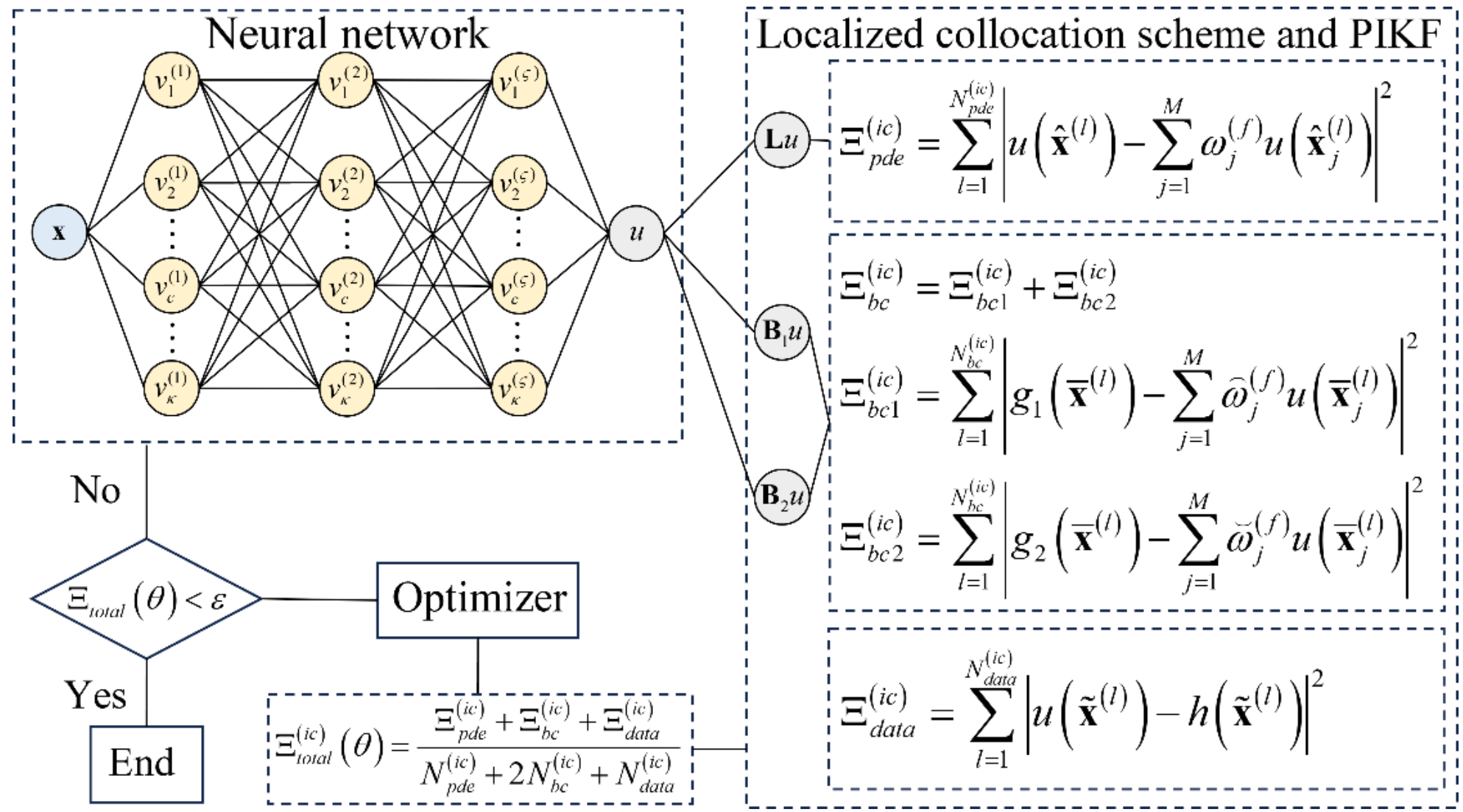

Fig. 4. Schematic diagram of the LPIKFNN inverse Cauchy problem framework.

## 3. Numerical examples

In this section, the first five examples are used to verify the accuracy and efficiency of the proposed LPIKFNN in simulating high wavenumber problems, higher-order derivative problems, nonlinear problems, and heterogeneous problems. The sixth example demonstrates the application of LPIKFNN to potential-based inverse electromyography of the pregnant sheep uterus. The relative error $\mathrm{rerr}$ and the $L_2$ relative error $\mathrm{Lrerr}$ are defined by

$$\mathrm{rerr}=\frac{\breve{u}\left(\mathbf{x}_i\right)-\widehat{u}\left(\mathbf{x}_i\right)}{\widehat{u}\left(\mathbf{x}_i\right)}, \tag{22}$$

$$\text{Lrerr} = \sum_{i=1}^{N_t} \left[ \breve{u}(\mathbf{x}_i) - \widehat{u}(\mathbf{x}_i) \right]^2 \Big/ \sum_{i=1}^{N_t} \left[ \widehat{u}(\mathbf{x}_i) \right]^2, \tag{23}$$

where $\breve{u}(\mathbf{x}_i)$ is the prediction value obtained by the LPIKFNN at the sample node $\mathbf{x}_i$. $\widehat{u}(\mathbf{x}_i)$ represents the analytical solution or experimental result, $N_t$ indicates the number of sample nodes. In this study, LPIKFNN and PINN are trained using the same configuration. Unless otherwise stated, the fully connected neural network consists of 5 hidden layers, each with 20 neurons. The number of training iterations is set to 5000, and the tanh activation function is employed.

### 3.1. Verification example

We first consider a wave propagation problem with Dirichlet boundary conditions to verify the accuracy and efficiency of the proposed LPIKFNN. A total of 4725 training nodes are distributed on the unit cube domain $\Omega = \{0 \le x_1, x_2, x_3 \le 1\}$, including 1350 boundary nodes and 3375 interior nodes, as depicted in Fig. 5. The governing equation and boundary condition for the wave propagation problem are given by

$$\Delta u(\mathbf{x}) + ku(\mathbf{x}) = 0, \mathbf{x} \in \Omega, \tag{24}$$

$$u(\mathbf{x}) = \sin\left( \frac{kx_1 + kx_2 + kx_3}{\sqrt{3}} \right), \mathbf{x} \in \Gamma, \tag{25}$$

where $\Delta$ denotes the Laplacian, $k$ is the wavenumber. The PIKF $F = \cos(kr)/4\pi r$ is incorporated into the LPIKFNN. A total of 1000 irregularly distributed nodes within the cubic domain are used as sampling nodes to evaluate the $L_2$ relative error, and wavenumber $k = 4$ is considered.

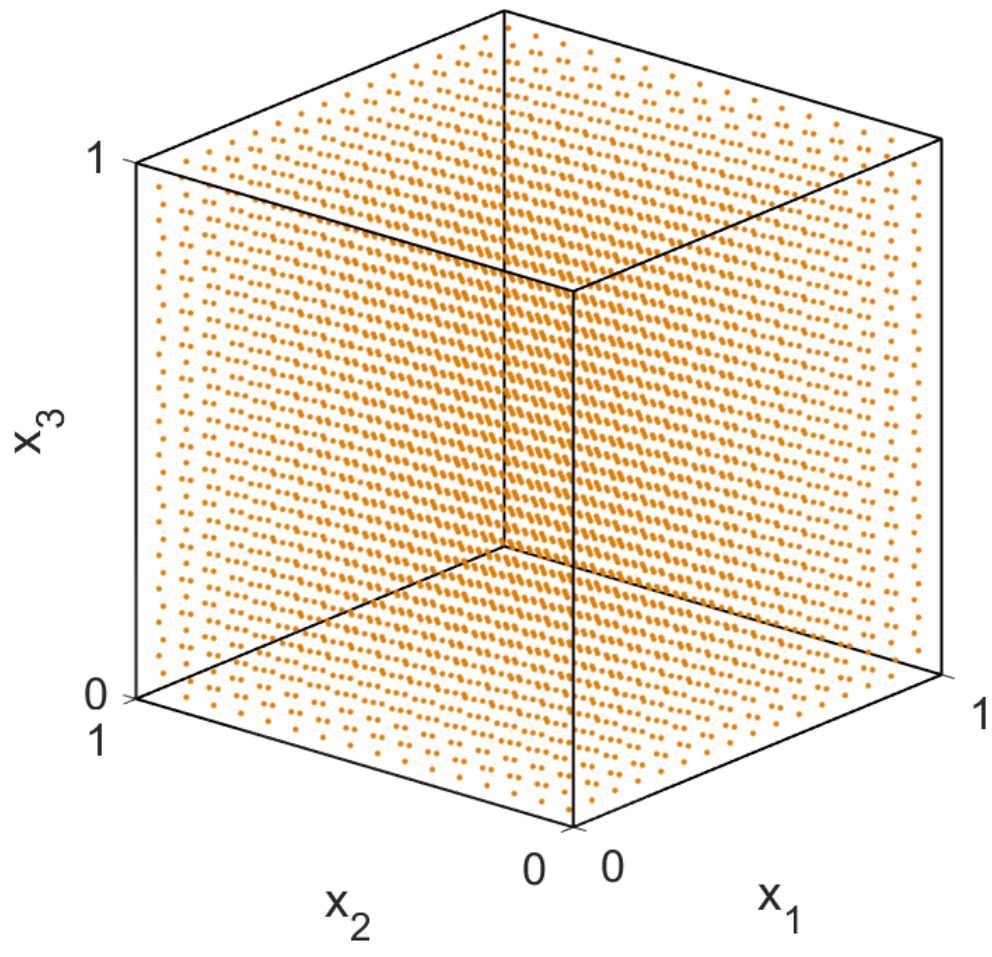

Fig. 5. Distribution of the training nodes.

We investigate the influence of the number of hidden layers and neurons on the prediction accuracy of LPIKFNN. Tables 1 and 2 lists the $L_2$ relative errors predicted by LPIKFNN with different number of hidden layers and neurons. It can be seen that there is little impact on the prediction accuracy with different number of hidden layers and neurons, the prediction results of LPIKFNN are accurate and stable, where the minimum $L_2$ relative error obtained at 5 hidden layers and 20 neurons. Then, the influence of different activation functions on prediction accuracy is considered. The $L_2$ relative errors obtained by LPIKFNN with four common activation functions are presented in Table 3. The prediction results clearly demonstrate that LPIKFNN achieves the highest prediction accuracy with tanh activation function. Therefore, the configuration in subsequent LPIKFNN predictions is as follows: 5 hidden layers, 20 neurons and tanh activation function.

Table 1 The $L_2$ relative errors with different number of hidden layers.

| Number of hidden layers | 3 | 4 | 5 | 6 | 7 |
|---|---|---|---|---|---|
| Lrerr | 2.68E-3 | 2.33E-3 | 2.18E-3 | 3.10E-4 | 3.61E-4 |

Table 2 The $L_2$ relative errors with different number of neurons.

| Number of neurons | 10 | 20 | 30 | 40 | 50 |
|---|---|---|---|---|---|
| Lrerr | 2.67E-3 | 2.18E-3 | 2.69E-3 | 3.62E-4 | 3.41E-4 |

Table 3 The $L_2$ relative errors with different activation functions.

| Activation function | tanh | ReLU | sigmoid | swish |
|---|---|---|---|---|
| Lrerr | 2.18E-4 | 4.48E-4 | 4.96E-4 | 3.44E-4 |

Next, we compare the prediction accuracy and training time of LPIKFNN, PINN, and localized method of fundamental solutions (LMFS). The $L_2$ relative errors and training times are summarized in Table 4. Lrerr represents the $L_2$ relative error of irregularly distributed nodes, and Lrerr1 represents the $L_2$ relative error of the training nodes. As observed, the prediction accuracy of LPIKFNN is approximately one order of magnitude higher than that of PINN and is comparable to that of LMFS. Moreover, the training time of LPIKFNN is approximately 40% of that of PINN. This improvement arises because LPIKFNN eliminates the need for automatic differentiation at each training iteration, thereby significantly enhancing training efficiency. It is worth noting that LMFS can only calculate the results at the training nodes, but cannot obtain the results at the irregularly distributed nodes. This is because LMFS lacks the training process of neural networks. By contrast, LPIKFNN and PINN can utilize pre-trained neural networks to generate prediction results for any irregularly distributed nodes within the computational domain.

Table 4 The $L_2$ relative errors and training times of LPIKFNN, PINN and LMFS.

| Method | LPIKFNN | PINN | LMFS |
| --- | --- | --- | --- |
| Lrerr | 2.02E-4 | 2.40E-3 | / |
| Lrerr1 | 2.18E-4 | 2.51E-3 | 2.09E-4 |
| Training time (s) | 298 | 751 | 2 |

### 3.2. High wavenumber problem

The second example is used to verify the performance of LPIKFNN in high wavenumber problems. Eqs. (24) and (25) are considered as governing equation and boundary condition. The computational domain and distribution of training nodes are the same as that in Example 1. Table 5 compares the prediction errors of LPIKFNN and PINN with different wavenumbers. The symbol "/" indicates that the $L_2$ relative error exceeds 1E-2. Notably, PINN fails to provide satisfactory prediction accuracy for wavenumbers above 7. In contrast, the presented LPIKFNN maintain high prediction accuracy even at wavenumbers up to 27, effectively overcoming the limitations of PINN in high wavenumber wave propagation simulations. Fig. 6 presents the relative error distributions on three different surfaces $\Gamma_i$, as predicted by LPIKFNN and PINN, where wavenumber $k=5$ and $\Gamma_i$ denotes the surface of the cubic domain $\Omega_i=\{i/10\le x_1,x_2,x_3\le 1-i/10\}, i=1,2,3$. It can be seen that the relative error of LPIKFNN is approximately one order of magnitude lower than that of PINN. The numerical results indicate that the presented LPIKFNN is reliable for predicting the high wavenumber problems.

Table 5 The $L_2$ relative errors with respect to wavenumber.

| k | LPIKFNN | PINN |
|---|---|---|
| 3 | 1.43E-4 | 7.01E-4 |
| 5 | 4.09E-4 | 3.06E-3 |
| 7 | 4.13E-4 | 4.21E-3 |
| 9 | 5.48E-4 | / |
| 11 | 7.81E-4 | / |
| 13 | 2.26E-3 | / |
| 15 | 2.94E-3 | / |
| 17 | 3.01E-3 | / |
| 19 | 3.21E-3 | / |
| 21 | 3.35E-3 | / |
| 23 | 4.88E-3 | / |
| 25 | 4.95E-3 | |
| 27 | 9.33E-3 | / |

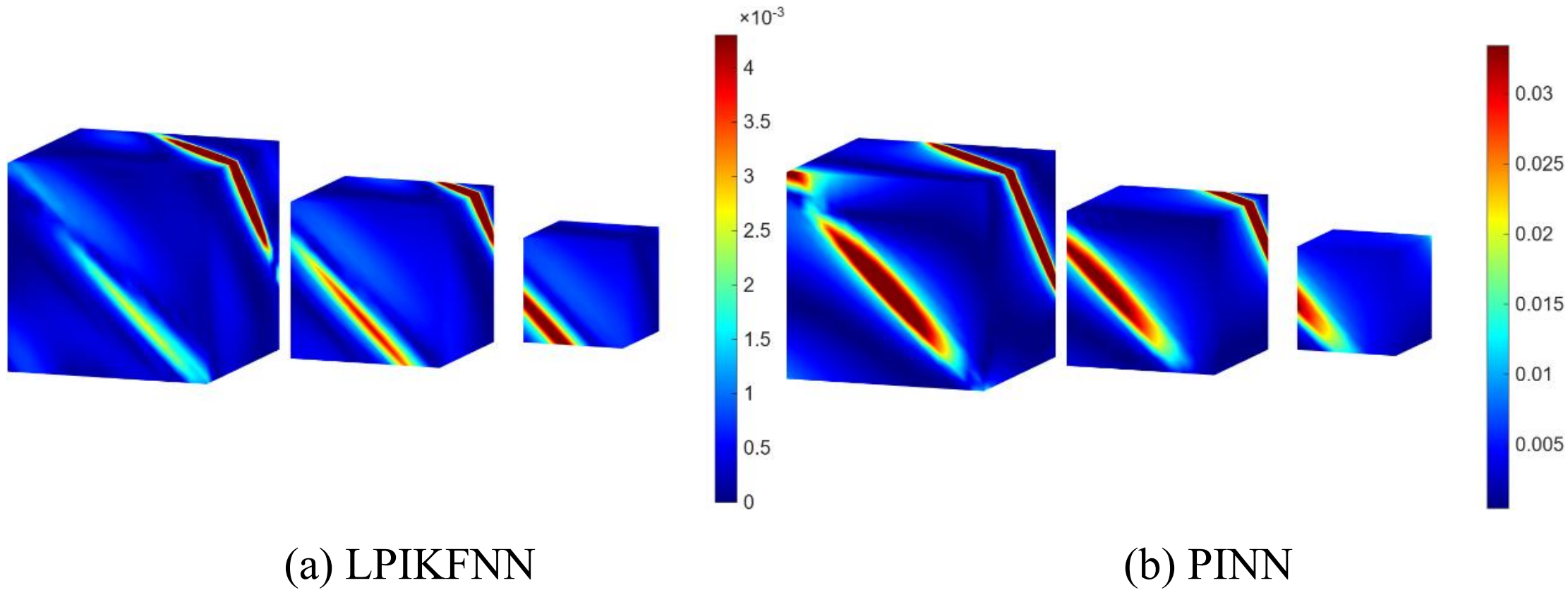

(a) LPIKFNN (b) PINN

Fig. 6. Relative error distributions on three different surfaces.

### 3.3. Higher-order derivative problem

The biharmonic problem is used to verify the performance of LPIKFNN in higher-order derivative. The computational domain and distribution of training nodes are the same as that in Example 1. The fourth-order governing equation and boundary conditions are given by

$$\Delta^2 u(\mathbf{x}) = 0, \mathbf{x} \in \Omega, \tag{26}$$

$$u(\mathbf{x}) = (x_1+1)^3 + (x_2+1)^3 + (x_3+1)^3, \mathbf{x} \in \Gamma, \tag{27}$$

$$\frac{\partial u(\mathbf{x})}{\partial \mathbf{n}} = 3(x_1+1)^2 n_{x_1} + 3(x_2+1)^2 n_{x_2} + 3(x_3+1)^2 n_{x_3}, \mathbf{x} \in \Gamma, \tag{28}$$

where $(n_{x_1}, n_{x_2}, n_{x_3})$ is the outward unit normal vector, Eqs. (27) and (28) represent Dirichlet boundary condition and Newman boundary condition, respectively. The PIKFs $F_1 = 1/4\pi r$ and $F_2 = r/8\pi$ are incorporated into the LPIKFNN.

To validate the effectiveness of LPIKFNN in simulating the biharmonic problem, Table 6 presents the comparison of the $L_2$ relative errors and training times between LPIKFNN and PINN. Fig. 7 provides the contours of relative errors on the three surfaces obtained by LPIKFNN and PINN. As shown in Table 6 and Fig. 7, the prediction accuracy of LPIKFNN for the biharmonic problem is approximately one order of magnitude higher than that of PINN, while its training time is only about 6% of that of PINN, thereby significantly reducing the overall computational cost.

Table 6 The $L_2$ relative errors and training times of LPIKFNN and PINN.

| Method | LPIKFNN | PINN |
|---|---|---|
| Lrerr | 2.10E-4 | 1.14E-3 |
| Training time (s) | 255 | 4226 |

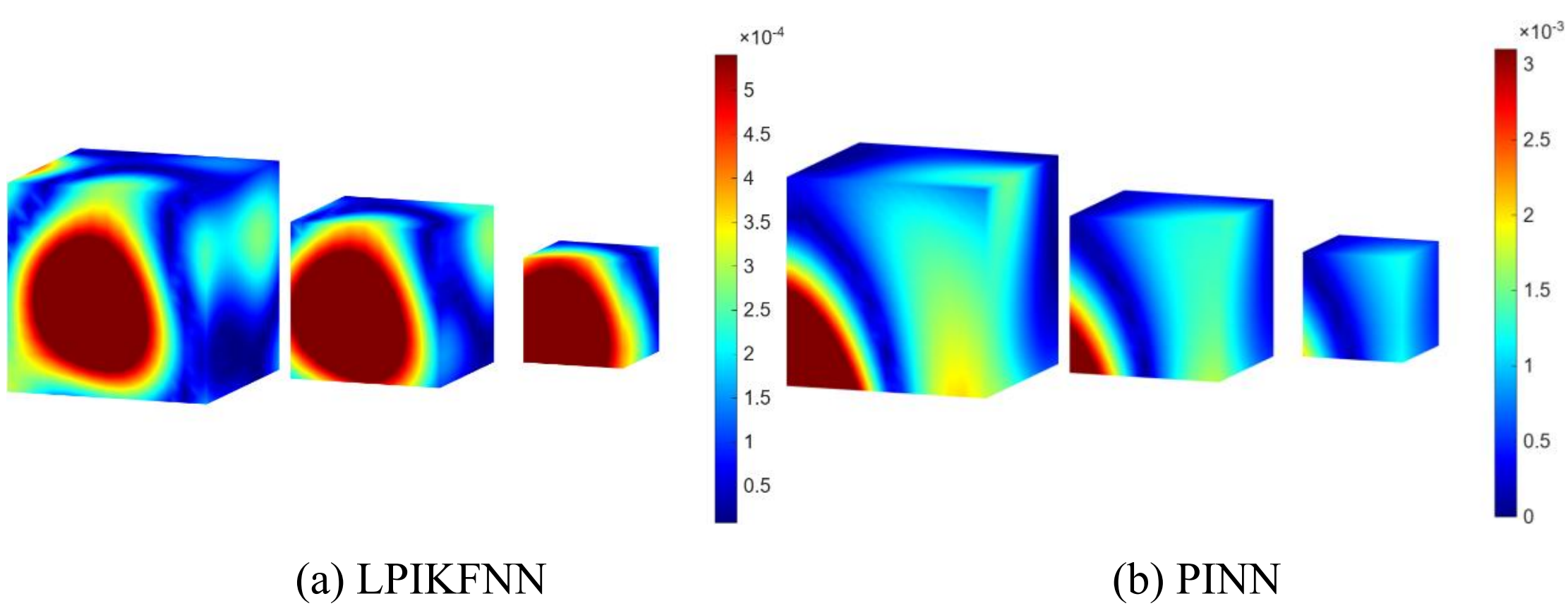

(a) LPIKFNN (b) PINN

Fig. 7. Contours of the relative errors on three different surfaces.

Furthermore, a comparison of the results in Tables 4 and 6 reveals a pronounced difference in training efficiency between LPIKFNN and PINN. Specifically, the training time of PINN for the biharmonic problem is approximately 5.6 times that required for the wave propagation problem, whereas LPIKFNN exhibits comparable training times in both cases. Notably, the biharmonic equation is a fourth-order PDE, while the wave propagation equation is second order. The numerical results indicate that the training cost of PINN increases significantly with the order of the PDE. In contrast, LPIKFNN demonstrates strong robustness with respect to equation order, maintaining stable training times for both second- and fourth-order problems, which remain substantially lower than those of PINN.

### 3.4. Nonlinear problem

It is noteworthy that identifying PIKFs that satisfy the global characteristics of nonlinear equations is generally challenging. This limitation is the primary reason LMFS cannot be applied directly to nonlinear problems. To overcome this challenge, the proposed LPIKFNN constructs the loss function using PIKFs that satisfy partial characteristics of the nonlinear equation and leverages the strong approximation capability of neural networks to achieve accurate predictions.

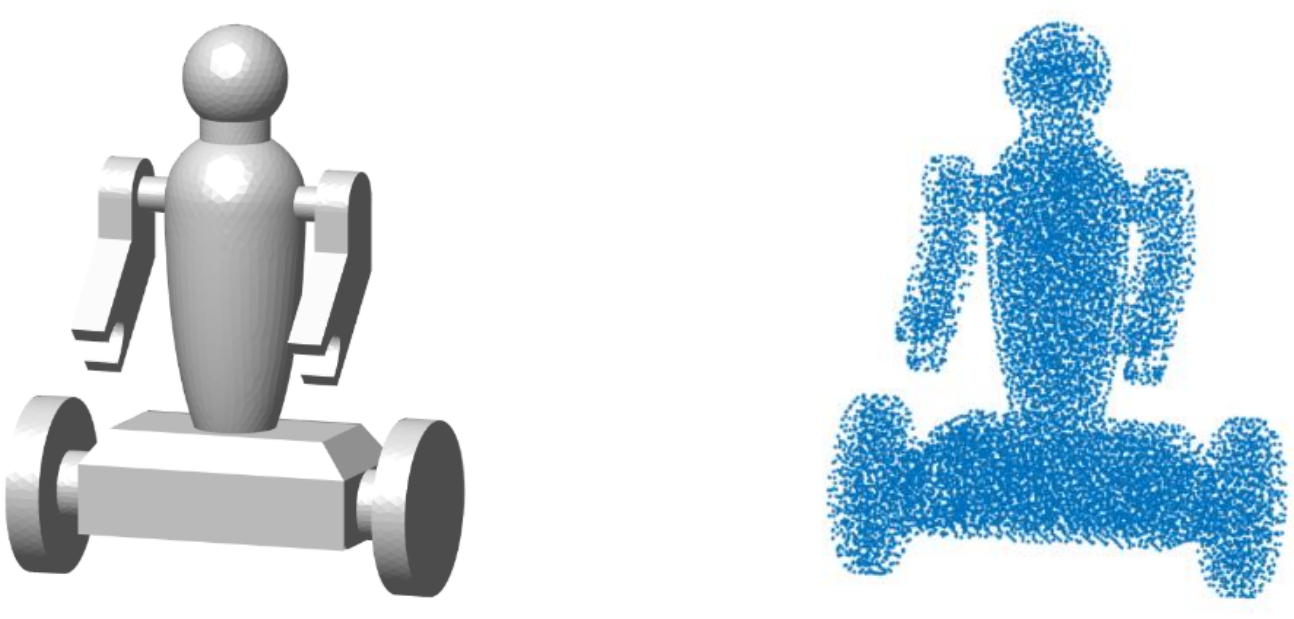

(a) Robot model (b) Distribution of the training nodes

Fig. 8. Schematic diagram of the robot model.

A nonlinear problem with Dirichlet boundary conditions defined on a complex robot model is considered. The dimension of the robot model as displayed in Fig. 8 is $0.5\text{m}\times 1.2\text{m}\times 1.6\text{m}$. The governing equation and boundary condition of the nonlinear problem are expressed as

$$\Delta u(\mathbf{x})-3u(\mathbf{x})-\frac{\partial u(\mathbf{x})}{\partial x_1}u(\mathbf{x})+u^2(\mathbf{x})=0,\mathbf{x}\in\Omega, \tag{29}$$

$$u(\mathbf{x})=e^{x_1+x_2+x_3},\mathbf{x}\in\Gamma. \tag{30}$$

During the training process of LPIKFNN and PINN, 4006 boundary training nodes are distributed on the surface of the robot model, while 9006 interior training nodes are distributed within the robot model. The PIKF $F = e^{-kr}/4\pi r$ is employed to construct the loss function in the LPIKFNN.

Table 7 presents the $L_2$ relative errors of $u$ predicted by LPIKFNN and PINN. It is evident that the prediction error of LPIKFNN is lower than that of PINN. This improvement arises because the PIKF employed in the LPIKFNN inherently encode physical information about the nonlinear problem, enabling the fully connected neural network to achieve higher accuracy with greater ease. The relative error distributions of $\partial u/\partial x_1$, $\partial u/\partial x_2$ and $\partial u/\partial x_3$ on the surface of the robot model predicted by LPIKFNN are depicted in Fig. 9. The numerical results clearly demonstrate that the LPIKFNN can achieve accurate prediction results for $\partial u/\partial x_1$, $\partial u/\partial x_2$ and $\partial u/\partial x_3$.

Table 7 The $L_2$ relative errors of $u$.

| Method | LPIKFNN | PINN |
| --- | --- | --- |
| *Lrerr* | 8.30E-5 | 3.15E-4 |

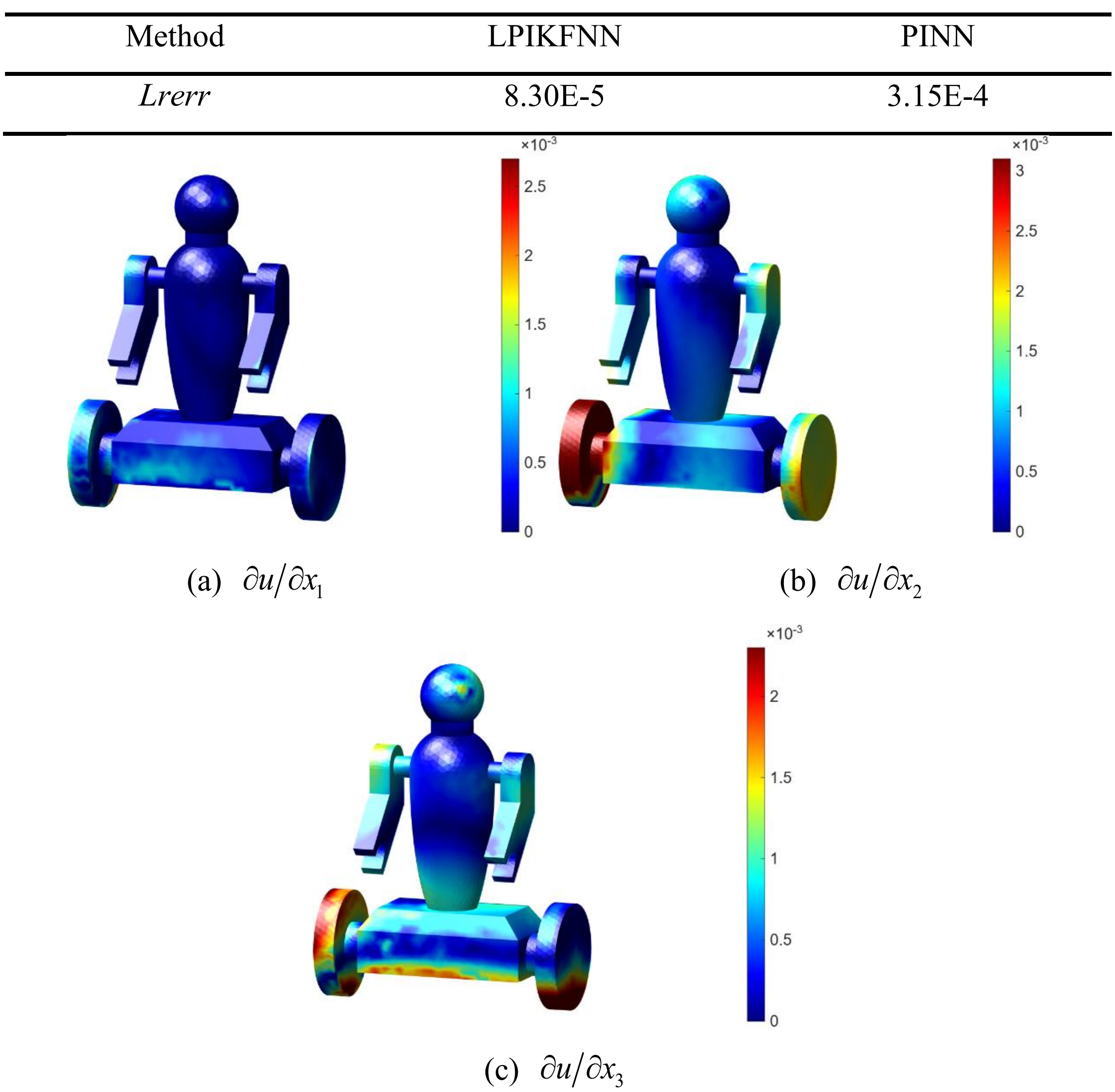

(a) $\partial u/\partial x_1$ (b) $\partial u/\partial x_2$

(c) $\partial u/\partial x_3$

Fig. 9. Relative error distributions on the robot surface predicted by LPIKFNN.

### 3.5. Heterogeneous problem

To verify that LPIKFNN eliminates the dependence of PIKFNN on the global PIKF, the heat conduction problem in a heterogeneous airplane model is considered. The dimension of the airplane model as displayed in Fig. 10 is $0.6\text{m}\times 0.2\text{m}\times 0.8\text{m}$. The governing equation and Dirichlet boundary condition of the heat conduction problem are given by

$$\Delta u(\mathbf{x})-\lambda(\mathbf{x})u(\mathbf{x})=0, \mathbf{x}\in\Omega, \tag{31}$$

$$u(\mathbf{x})=x_1e^{x_1}+x_2e^{x_2}+x_3e^{x_3}, \mathbf{x}\in\Gamma. \tag{32}$$

where $\lambda(\mathbf{x})=1+\dfrac{2\left(e^{x_1}+e^{x_2}+e^{x_3}\right)}{x_1e^{x_1}+x_2e^{x_2}+x_3e^{x_3}}$ represents heterogeneous parameter. It should be noted that $\lambda(\mathbf{x})$ is not constant, and the global PIKF associated with the heterogeneous parameter $\lambda(\mathbf{x})$ is unavailable in the PIKFNN, thereby preventing it from directly predicting this heterogeneous heat conduction problem. In contrast, the heterogeneous PIKF $F(\mathbf{x})=e^{-\lambda(\mathbf{x})r}/4\pi r$ is employed in LPIKFNN to construct the loss function for neural network training.

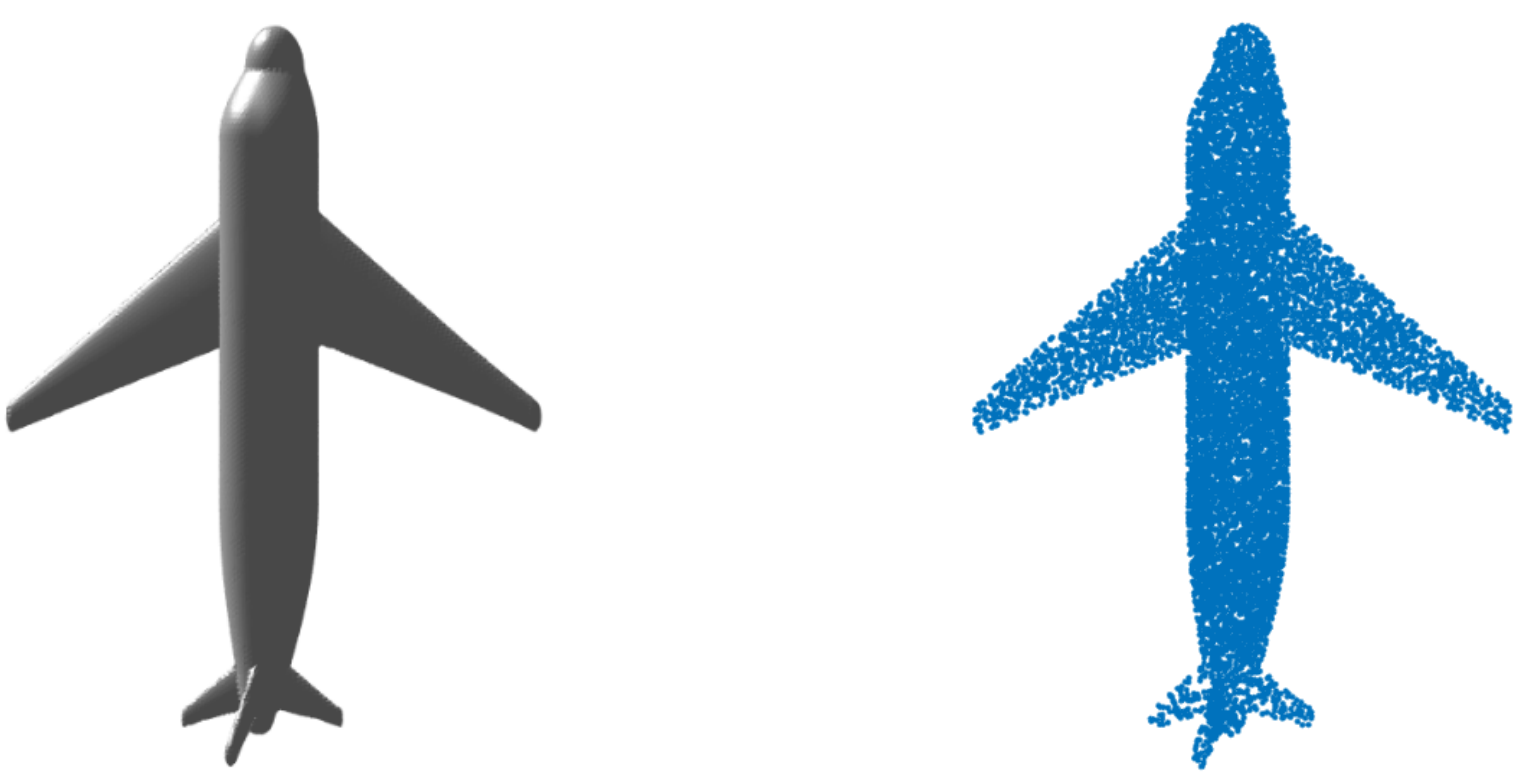

(a) Airplane model (b) Distribution of the training nodes

Fig. 10. Schematic diagram of the airplane model.

During the training process, 3369 boundary nodes are distributed over the surface of the airplane model, while 5625 interior nodes are located within the airplane model. The $L_2$ relative errors predicted by LPIKFNN and PINN are listed in Table 8. The results indicate that the prediction accuracy of the temperature obtained by LPIKFNN is higher than that of PINN. Fig. 11 presents the contours of the relative errors for the heat flux components on the surface of the airplane model. It can be seen that the LPIKFNN achieves satisfactory prediction accuracy for all heat flux components. These numerical results confirm that the LPIKFNN is both effective and accurate, highlighting its strong potential for analyzing heterogeneous heat conduction problems.

Table 8 The $L_2$ relative error of temperature $u$.

| Method | LPIKFNN | PINN |
|---|---|---|
| *Lrerr* | 6.86E-5 | 4.37E-4 |

(a) $\partial u/\partial x_1$ (b) $\partial u/\partial x_2$

(c) $\partial u/\partial x_3$

Fig. 11. Relative error distributions on the airplane surface obtained by LPIKFNN.

### 3.6. Inverse electromyography problem

The final example demonstrates the application of the proposed LPIKFNN in potential-based inverse electromyography of the pregnant sheep uterus by comparing it with experimental data. Under the electro-quasi-static assumption, the electric potential measured on the body and uterus of the pregnant sheep at each time frame satisfies the Laplace equation. The surface of the sheep's body is assumed to be an insulating medium, and the electric flux on the body surface is set to zero, corresponding to a Neumann boundary condition. The electrical potential data measured experimentally on the body surface at each time frame are imposed as Dirichlet boundary conditions. Accordingly, the governing equation and boundary conditions for the potential-based inverse electromyography problem can be expressed as

$$\Delta u(\mathbf{x}) = 0, \mathbf{x} \in \Omega, \tag{31}$$

$$\frac{\partial u(\mathbf{x})}{\partial \mathbf{n}} = 0, \mathbf{x} \in \Gamma_m, \tag{32}$$

$$u(\mathbf{x}) = \hat{u}_{data}(\mathbf{x}), \mathbf{x} \in \Gamma_m, \tag{33}$$

where $\hat{u}_{data}(\mathbf{x})$ represents the measured electrical potential data, $\Gamma_m$ is the surface of the pregnant sheep's body. This problem is a typical inverse Cauchy problem. A schematic diagram of the pregnant sheep's body and uterus is depicted in Fig. 12. The LPIKFNN is employed to construct the loss function associated with the governing equation, boundary conditions, and measurement data, and to predict the electrical potential on the uterine surface at each time frame.

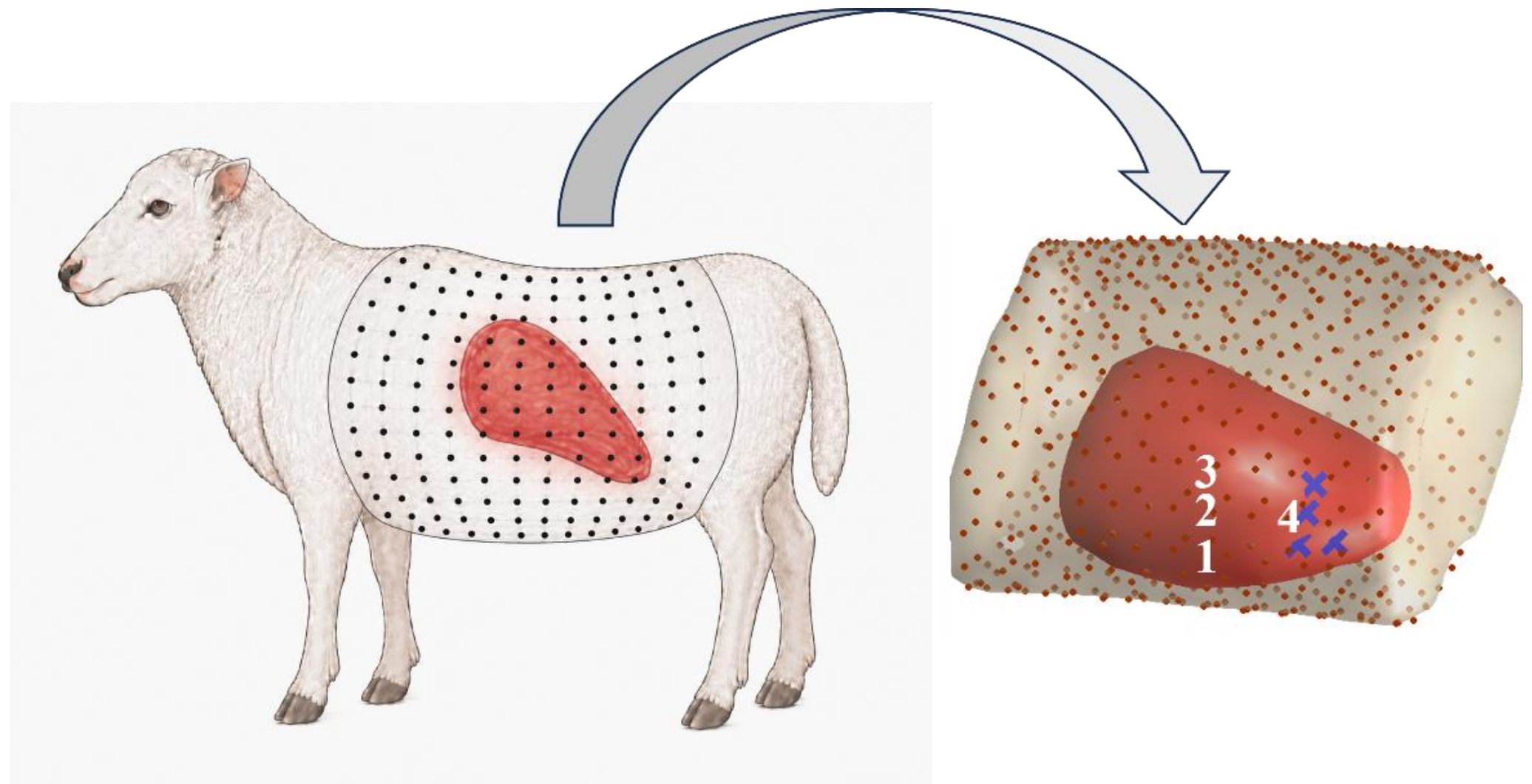

Fig. 12. Schematic diagram of the pregnant sheep's body and uterus.

The electric potentials on the body and uterus of the pregnant sheep are measured using electrodes placed on their surfaces. The experiment with the pregnant sheep lasted for 223.2 seconds, and the electric potential data were recorded every 0.01 seconds. During the LPIKFNN training process, 707 outer boundary training nodes are distributed on the body surface, 114 inner boundary training nodes are on the uterine surface, and 997 interior training nodes are placed within the domain between the body and the uterus. The PIKF $F = 1/4\pi r$ is used to construct the loss function. To evaluate the agreement and error between the predicted electrical potentials and the experimental data, the correlation coefficient (CC) and the relative error (RE) are defined by

$$\text{CC} = \sum_{t=1}^{T}\left(u_t^{(p)} - \bar{u}^{(p)}\right)\left(u_t^{(e)} - \bar{u}^{(e)}\right) \Bigg/ \sqrt{\sum_{t=1}^{T}\left(u_t^{(p)} - \bar{u}^{(p)}\right)^2 \sum_{t=1}^{T}\left(u_t^{(e)} - \bar{u}^{(e)}\right)^2}, \tag{34}$$

$$\text{RE} = \sum_{t=1}^{T}\left(u_t^{(p)} - u_t^{(e)}\right)^2 \Bigg/ \sum_{t=1}^{T}\left(u_t^{(e)}\right)^2, \tag{35}$$

where $T$ is the number of time intervals, $u_t^{(p)}$ represents the predicted electrical potential value, $u_t^{(e)}$ denotes the experimentally measured electrical potential value, $\bar{u}^{(p)}$ and $\bar{u}^{(e)}$ are the average values of the predicted and experimentally measured electrical potentials.

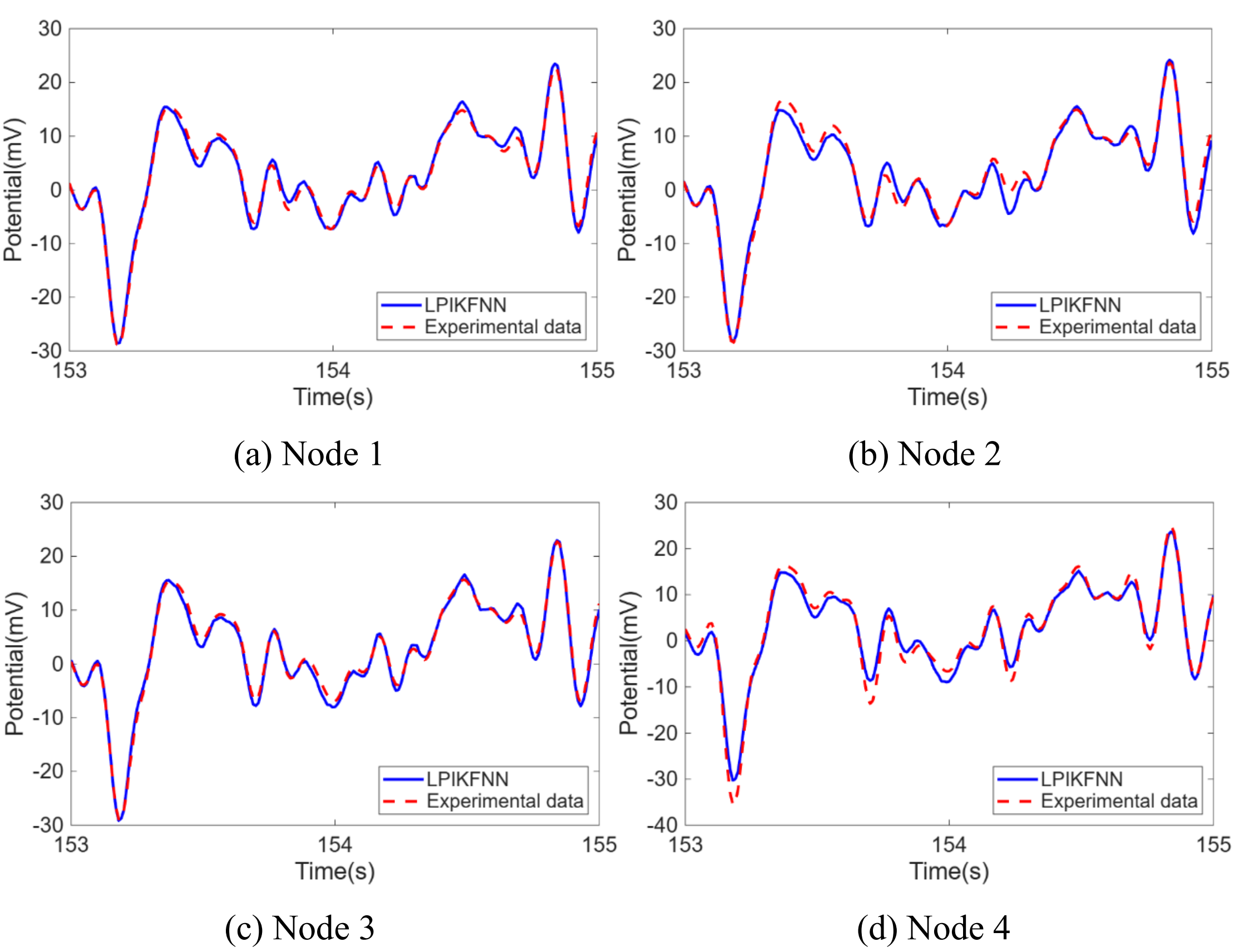

Fig. 13. The electrical potential obtained by LPIKFNN and experimental data.

Fig. 13 compares the electrical potentials predicted by LPIKFNN and the experimental data. The locations of the four electrodes on the uterine surface are shown in Fig. 13. The experimental measurement time ranges from 153 seconds to 155 seconds, corresponding to $T = 201$. The numerical results demonstrate that the electrical potentials predicted by LPIKFNN are in good agreement with the experimental data. The correlation coefficients and relative errors between the predicted values of LPIKFNN and PINN and the experimental data are listed in Table 9. It can be observed that the correlation coefficient of the LPIKFNN predictions is higher than that of the PINN predictions, with both exceeding 99%. Furthermore, the relative error of the LPIKFNN predictions is lower than that of the PINN predictions, which is an acceptable accuracy for clinical applications [44]. These results verify that the proposed LPIKFNN provides accurate predictions and can serve as an alternative tool for potential-based inverse electromyography.

Table 9 The CC and RE determined by LPIKFNN and PINN.

| | | Node 1 | Node 2 | Node 3 | Node 4 |
|---|---|---|---|---|---|
| LPIKFNN | CC | 99.64% | 99.55% | 99.07% | 99.01% |
| | RE | 8.52E-2 | 9.15E-2 | 1.37E-1 | 1.59E-1 |
| PINN | CC | 99.62% | 99.51% | 98.85% | 98.46% |
| | RE | 8.86E-2 | 1.04E-1 | 1.77E-1 | 2.21E-1 |

## 4. Conclusions

This paper presents a localized physics-informed kernel function neural network (LPIKFNN) for predicting steady-state physics problems. In contrast to the conventional PINN, LPIKFNN constructs the loss function using the PIKF and localized collocation scheme rather than relying on automatic differentiation, thereby eliminating the costly derivative evaluations required to enforce the governing equations during iterative training. The effectiveness of the LPIKFNN is validated through comparisons with analytical solutions, PINN predictions, and experimental results. The main conclusions of this study are summarized as follows.

- LPIKFNN eliminates the need for automatic differentiation during iterative training in conventional PINNs, thereby reducing model training time. Specifically, its training time for the fourth-order PDE is only about 6% of that of PINN.
- LPIKFNN incorporates PIKF into the loss function, significantly enhancing the computational accuracy for high-wavenumber problems.

- LPIKFNN overcomes the computational bottleneck of LMFS and achieves accurate prediction of nonlinear problems.
- LPIKFNN eliminates the reliance on global PIKF in PIKFNN by introducing the localized collocation scheme, thereby enabling accurate predictions for heterogeneous problems where a global PIKF is unavailable.
- LPIKFNN prediction results show good agreement with experimental results in the potential-based inverse electromyography of the pregnant sheep uterus.

Finally, we present a summary of LPIKFNN, PINN, LMFS, and PIKFNN in Table 10 to better understand the characteristics of PIKFNN. In summary, LPIKFNN is an enhanced PINN framework that achieves higher accuracy and efficiency than conventional PINN for physical problems where PIKFs are available. However, it is not applicable to problems for which PIKFs are unavailable. This limitation is still under investigation and will be reported in subsequent papers.

Table 10 Summary of LPIKFNN, PINN, LMFS, and PIKFNN [a)]

| Problems | LPIKFNN | PINN | LMFS | PIKFNN |
|---|---|---|---|---|
| High wavenumber problems | **** | ** | **** | ***** |
| Higher-order derivative problems | **** | ** | **** | **** |
| Nonlinear problems | *** | **** | * | ** |
| Heterogeneous problems | ***** | **** | ***** | ** |
| Scope of application | **** | ***** | *** | *** |
| Interpretability | *** | * | *** | **** |

a) more "*" much better in the comparison.

**Acknowledgements**

The work reported in this paper was supported by the Natural Science Foundation of China (Grant Nos. 12302258, 12372196), the Fundamental Research Funds for the Central Universities (Grant No. B250201051).